# Radiation Damage and Thermal Recovery of Perovskite Superconductor Yttrium Barium Copper Oxide

A. Khobnya[1], M. E. Pek[1], G. Greaves[2], M. Danaie[3], G. D. Brittles[4], S. E. Donnelly[2], F. Schoofs[5], A. Reilly[5], P. D. Edmondson[6], S. Pedrazzini[1]

[1] Department of Materials, Imperial College London, Exhibition Road, SW7 2AZ, London, UK.
[2] School of Computing and Engineering, University of Huddersfield, Queensgate, Huddersfield HD1 3DH, UK.
[3] Diamond Light Source, electron Physical Science Imaging Centre (ePSIC), Harwell Science and Innovation Campus, Fermi Ave, Didcot OX11 0DE, UK.
[4] Department of Materials, University of Oxford, Parks Road, OX1 3PH, Oxford, UK.
[5] UK Atomic Energy Authority, Culham Science Centre Abingdon OX14 3EB, UK.
[6] Oak Ridge National Laboratory, 1 Bethel Valley Road, Oak Ridge, TN 37831, Tennessee, USA.

**Abstract**

High-temperature superconducting materials are being considered to generate the magnetic fields required for the confinement of plasma in fusion reactors. The present study aims to assess the microstructural degradation resulting from ion implantation at room temperature under two implantation conditions: 0.6 MeV $Xe^{2+}$ to a fluence of $9 \times 10^{13}$ ions/cm$^2$ and 2 MeV $Xe^{2+}$ ions to a fluence of $1 \times 10^{16}$ ions/cm$^2$, in Yttrium Barium Copper Oxide (YBCO) tapes. X-ray Diffraction (XRD) and high-resolution characterisation techniques including Transmission Electron Microscopy (TEM) analyses were used to correlate alterations in superconducting behaviour measured using a Magnetic Properties Measurement System (MPMS) to amorphization and recovery caused by ion implantation. TEM analysis was performed to depth-profile the degree of crystallinity (or lack thereof) on irradiated samples. SRIM predicted the damage depth at 900 nm below the sample surface of the 2 MeV $Xe^+$ implanted sample and 450 nm beneath the surface of the 0.6 MeV $Xe^{2+}$ implanted sample. 2 MeV $Xe^+$ implantation caused the superconducting temperature to decrease by 10 K and the critical current density to display a 10-fold reduction.

Post-irradiation heat treatments up to 600°C caused recrystallisation of the irradiated layer, but also oxygen loss and alterations in grain size. The recrystallised grain orientation was random in TEM lamellae, however, bulk samples re-grew along the original crystal orientation provided that some of the original material was not amorphized (ie if they nucleated on crystalline YBCO). This is promising for the thermal recovery of tokamak components.

### 1. Introduction:

The high temperature superconductors (HTS) based on Yttrium Barium Copper Oxide (YBCO) and other rare-earth elements (collectively, REBCO) have been proposed as candidate materials for magnetic confinement of plasma in Tokamak nuclear fusion reactors (*1*). Knowledge of the conditions which will cause degradation of the superconducting properties to a level at which they will no longer create confinement is a crucial consideration, which can then inform the reactor design so that an adequate level of shielding can be selected in order to make reactor operation commercially viable.

It is therefore essential, for the assessment of candidate materials, to study their electrical and microstructural response to radiation damage. Fischer et al. (*2*) irradiated various YBCO and REBCO tapes of different brands to fast neutron fluences of up to $3.9 \times 10^{22}$m$^{-2}$. They examined the critical current ($I_c$) up to fields of 15 T and down to 30 K. They found an initial increase in $I_c$, peaking at different fluences depending on sample type and whether or not they had artificial pinning centers (APCs). Tapes with APCs degraded at lower fluences than tapes without them. Artificial pinning centres can be, for example, $BaZrO_3$ self-assembling columns or other columnar defects, including amorphous ion-induced cascades in the otherwise crystalline lattice (*3*).
Ion implantation is often used instead of neutron irradiation to simulate the damage without activating the samples, and the extensive literature on the validity and limitations of this approach can be found elsewhere (*4*, *5*). Room temperature ion implantation to low-fluences has been shown to increase the critical current density of YBCO through generation of new pinning sites (*6*). This effect is counterbalanced at higher fluences by damage within the crystal structure and oxygen loss, both of which can arise from radiation damage and reduce the critical current density carried by the superconductor (*2*). The exact transition point after which the beneficial effects become detrimental is not well understood (though some studies are starting to assess it for a number of coated conductors (*2*)) and is a key limiting factor in designing the potential lifetime of the superconducting windings (*7*, *8*).

Previous studies on irradiated YBCO characterised the implanted material, but only at relatively low fluences, up to $1 \times 10^{13}$ ions/cm$^2$ (*6*, *9–14*). These studies, on neutron or ion irradiated YBCO focus on the beneficial effects, which include small increases in critical temperature $T_c$, and comparatively large increases in critical current density $J_c$ through the generation of new flux pinning sites (*9–13*). Few studies found where a transition point beyond which the effects become detrimental was mapped: Suvorova *et al.* (*6*) irradiated samples of SuperPower tape with 107 MeV Kr$^+$ ions varying the fluence between $2 \times 10^{10}$ and $6 \times 10^{13}$ ions/cm$^2$. They found that above a fluence of $8 \times 10^{11}$ ions/cm$^2$ both $J_c$ and $T_c$ decreased abruptly, however, they only characterised the sample in which the superconducting properties were maximised. In another study, Antonova *et al.* (*14*) used SuperPower HTS tape samples, in which the Cu stabilising layer was etched off revealing the Ag overlayer, which was not removed. They irradiated their samples with varying fluences of 167 MeV Xe$^+$, 107 MeV Kr$^+$ and 48 MeV Ar$^+$ ions and found that the critical temperature $T_c$ starts to reduce at fluences $\phi \geq 1 \times 10^{11}$ ions/cm$^2$ and that above $5 \times 10^{12}$ ions/cm$^2$ superconductivity is completely absent. They complemented these measurements with X-ray diffraction, with which they proved that the material was fully amorphous above those fluences. They noted that diffraction peak positions remained unaltered with increasing fluence, only their intensity decreased. This indicated that no lattice swelling occurred due to ion implantation, only amorphisation.

While an understanding of the amorphisation conditions of YBCO is essential to its safe use as a magnetic confinement material for nuclear applications, the concept of post-amorphisation thermally induced recrystallisation is also worthy of investigation(*7*). This method would potentially allow component lifetime extension and therefore a simultaneous reduction in radioactive waste. The superconducting tape is coated in copper, therefore electro-thermal treatment of the copper could allow the recovery of the underlying superconducting tape. This concept has been partially tested by Tate *et al.* (*15*), who showed that rapid thermal annealing (20 s at 1143 K in oxygen) could recrystallise the samples and recover the properties.

There are two essential aspects which must be addressed by the characterisation procedure: accurate atomic-scale quantification of the oxygen stoichiometry and an assessment of the level of amorphisation/crystallinity, both of which can be altered by exposure to radiation and both of which are crucial for superconducting properties. The present study relies on laboratory X-ray diffraction (XRD) as a method which allows the measurement of lattice parameters (which can then be correlated to oxygen stoichiometry), and transmission electron microscopy (TEM) to assess the degree of crystallinity/amorphisation caused by exposure to ion implantation. The two techniques combined allow a comprehensive characterisation of the microstructures which, complemented by the use of superconducting quantum interference device (SQUID) magnetometry measurements, leads to improved understanding of the degradation of superconducting properties due to ion implantation.

2. **Experimental Methods:**
<u>2.1 Sample production:</u>
Single crystal YBCO samples were produced by the Top Seeded Melt Growth (TSMG) process (*16*) by the Bulk Superconductivity Group at the University of Cambridge. Powders of the required compositions were pressed into a compact pellet, followed by melting peritectic regrowth of the $Y_1Ba_2Cu_3O_{7-\delta}$ phase (Y-123). A chemically stable $GdBa_2Cu_3O_{7-\delta}$ seed was used as a heterogeneous nucleation site for epitaxial growth of a large Y-123 single crystal, containing additional $Y_2Ba_1Cu_1O_5$ phase (Y-211) particles as inclusions. An optimum amount of 20-30% of $Y_2Ba_1Cu_1O_5$ phase is required to produce an microstructure for improved flux pinning (*16*). Vertical sections (a-c plane) were prepared by cutting along the <a> axis and polishing to a 1μm diamond finish using ethanol as a lubricant.

Samples of 4 mm wide 2G-HTS tape from SuperPower Inc (with APCs) were also analysed and irradiated for comparison with the bulk irradiated samples and for ease of superconductivity measurements. The 1μm thick layer of YBCO, which was delivered with a Cu stabilising layer and Ag overlayer was exposed by etching with ferric chloride and a combination of ammonium hydroxide and hydrogen peroxide respectively. The oxide-based buffer layers and Hastelloy substrate on which the YBCO was grown were however not removed, maintaining the structural integrity of the tape. Previous measurements have shown that this chemical etching process does not harm the superconducting properties of the wire (*17*).

### 2.2 Magnetic properties measurements:
The samples of Superpower tape, with HTS surface exposed, were cut into 3 mm diameter discs, mounted in standard plastic straws and loaded into the MPMS with the applied field held perpendicular to the surface of the tape. Superconductivity temperature and critical current density measurements were performed using a Quantum Design SQUID, using the 3 mm disc samples of etched SuperPower tape, oriented with the applied field B//<c> axis. Measurements were performed varying the applied field from 0.01 – 6 T and the temperature from 4.2 – 77 K.

### 2.3 Implantation depth and damage simulations:
SRIM (*18*) was used to simulate the implantation profiles of the bulk and tape samples, as well as the in-situ ion implanted TEM 3mm disc (prepared from the Superpower tape). The "Ion Distribution and Quick Calculation of Damage" (Kinchin-Pease) model was used with bulk density set to 5.51 g/cm$^3$, atomic density to 7.65x10$^{22}$ atoms/cm$^3$, displacement energy of 25 eV for yttrium, barium and copper and 20 eV for oxygen. The ion beam conditions used were 2 MeV Xe for the ex-situ and 600 keV Xe for the in-situ irradiations.

### 2.4 Ion implantation experiments:
Ex-situ ion implantations were performed on YBCO bulk and tape samples along the crystallographic <c> axis at the University of Surrey Ion Beam Centre. 2 MeV Xe$^+$ ions were implanted at room temperature into the samples to a total fluence of 1x10$^{16}$ ions/cm$^2$. A summary of the change in superconducting critical temperature following the ex-situ implantations is shown in Table 1. To obtain a fluence for amorphisation, in-situ implantations were performed within a TEM using the MIAMI (Microscopes and Ion Accelerators for Materials Investigations) facility at the University of Huddersfield. In-situ ion implantations were performed on electron transparent sections of 3mm TEM discs (preparation method detailed below) at room temperature, utilising the MIAMI-2 system, which consists of a Hitachi 9500 TEM (operated at 300keV) combined with two ion beamlines, details of which are given elsewhere (*19*). These samples were irradiated along the <c> axis with 600 keV Xe$^{2+}$ at a flux of 6x10$^{12}$ ions/cm$^2$/s to a total fluence of 9x10$^{13}$ ions/cm$^2$. Prior to irradiation, electron-energy loss spectroscopy (EELS) was performed to confirm the starting composition of the area under investigation. The irradiation was performed at room temperature, with the irradiation periodically stopped to record images and selected area diffraction patterns. The sample was observed with the electron beam during ion implantation. Following amorphisation, the sample was heated in 100°C increments up to 900°C, using a Gatan 652 heating holder, to observe recrystallization of the YBCO. Azimuthal integrations of the diffraction patterns were plotted and used to assess the degree of crystallinity of samples.

### 2.5 Characterisation:
Samples were prepared for TEM analyses using standard lift-out techniques on a Zeiss NVision 40 focussed ion beam (FIB) microscope. TEM samples were prepared for the in-situ ion implantation and recovery study, by cutting 3 mm discs from the etched Superpower tape using a 3mm disc cutter. The discs were mounted using a Gatan clamp-style DuoPost holder, with the perovskite side facing downwards. A Gatan PIPS II ion polishing system was used to ion-beam mill the samples until electron transparency was obtained. The ion guns were set to 6˚ from the top (both), the beam energy was set to 5 keV and the samples were spun at 3.5 revolutions/min. Double ion-beam modulation was used and beam currents were in the range 20-30 μA. TEM analysis was performed on a JEOL 2100 microscope equipped with an Oxford Instruments energy dispersive X-ray spectroscopy (EDX) detector. Crystallinity and amorphisation were shown through the acquisition of selected area diffraction patterns across the ion implanted area.

A Bruker D2 Phaser (equipped with a 1D strip detector) was used for X-ray diffraction (XRD) scans with a voltage of 30 kV, a current of 10 mA using variable step size, angle and dwell time which will be specified on each scan, on a stage spinning at 60 revolutions/min. Bulk samples were mounted with plasticine onto a polycarbonate sample holder. SuperPower tape was placed on level single crystal silicon. A glass slide was used to level the samples and ensure they were parallel to the rim of the sample holder. Commercial software CrystalDiffract was used to analyse the patterns along with entries from the ICSD database (© FIZ Karlsruhe). Two methods for the calculation of lattice parameters were compared: (1) using XRD patterns from the ICDD database which were performed on compounds of known oxygen content and finding the best match and (2) using a pattern for fully oxygenated Y-123 perovskite and adjusting the lattice parameter until the best fit was found, then using the difference in c-axis parameter to calculate the oxygen content according to the formula developed by Benzi et al (*20*). Ultimately, since it was difficult to establish how the data was collected for each paper from the ICDD database, the paper by Benzi et al (*20*) was used, because we could at least ensure that the measurements were performed on the same type of sample, prepared in the same way, therefore lattice parameter changes could not be a result changes in sample preparation or in substitutional rare earths within the YBCO lattice.

## 2.6 Ex-situ Thermal Recovery:

Heat treatments were carried out using the Carbolite Gero Rapid Heating Chamber Furnace RWF in air at atmospheric pressure. The furnace was heated 15˚C above the temperature needed for annealing. Once the stable temperature was reached, samples were placed into the furnace. Opening the furnace dropped the temperature by around 15˚C (to the intended temperature of annealing), once closed, the target temperature was reduced by 15˚C to maintain the annealing temperature. After the samples were held at the annealing temperature for the intended time, they were removed from the furnace and allowed to air cool to room temperature.

### 3. Results:

#### 3.1: Microstructural characterisation of the unirradiated samples

The TSMG process produced a homogeneous microstructure consisting of discrete Y-211 particles <5 μm in diameter within the single crystal Y-123 matrix, as shown in our previous published work on experimental setting optimisation for YBCO characterisation (*21*).

X-ray diffractograms of the pristine (non-irradiated) YBCO melt-grown single crystal (SX) bulk sample confirmed that the $YBa_2Cu_3O_{7-\delta}$ (Y-123) and $Y_2BaCuO_5$ (Y-211) phases were present (Figure 1). In the Superpower tape, the presence of Y-123, γ nickel and MgO phases were confirmed (Figure 1). The data collected by XRD was used to determine <c>-axis lattice parameters of the Y-123 phase in the bulk and tape, which were 11.7022 Å and 11.7243 Å respectively. The Y-123 phase in both types of sample was confirmed to have preferential orientation along the <c> axis, as expected due to the production methods.

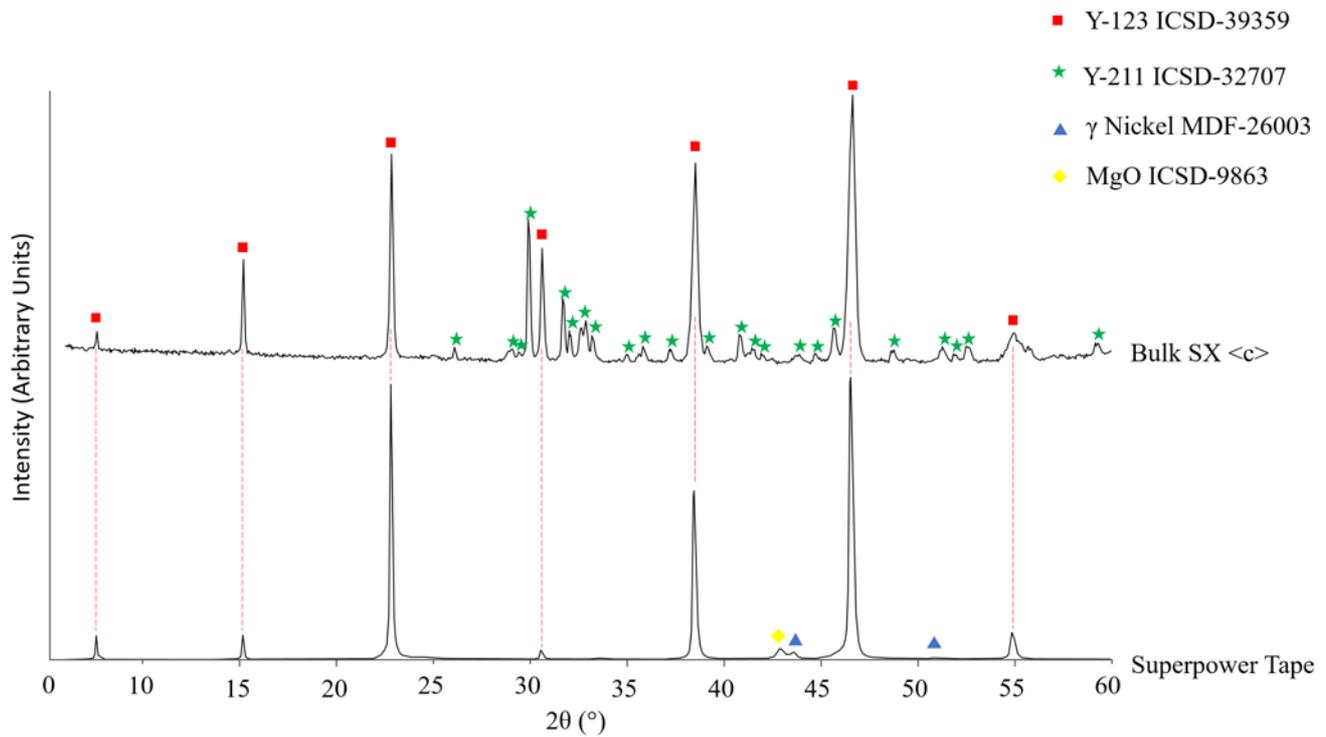

*Figure 1*: X-ray Diffractograms of pristine bulk single crystal YBCO, oriented on the <c> axis, and Superpower tape, etched to reveal the YBCO layer. Indicating the presence of 2 phases (Y-123 and Y-211) in the bulk sample and 3 phases (Y-123, γ Ni and MgO) in the tape.

### 3.2: SRIM/TRIM theoretical predictions of implantation depths and damage

SRIM (Stopping Ranges of Ions in Matter) and TRIM (TRajectories of Ions in Matter), as open access resources, have become common practice to predict the damage depths and calculate doses produced by ion implantation (*22*). Sources of errors can arise as SRIM assumes an amorphous target and relies on a Monte Carlo approach using the binary collision approximation. Nevertheless, the predicted damage layer depth of the ex-situ ion implantation performed at the Surrey Ion Beam Centre (Figure 2.a) was around 800 nm, with peak damage (40 dpa) around 300 nm, when implanted with 2 MeV $Xe^+$ to a fluence of $1\times10^{16}$ ions/cm$^2$. The predicted damage layer depth of the in-situ ion-irradiated samples (Figure 2.b) was around 300 nm, with peak damage (0.45 dpa) around 100 nm when implanted with 600 keV $Xe^{2+}$ to a fluence of $9\times10^{13}$ ions/cm$^2$.

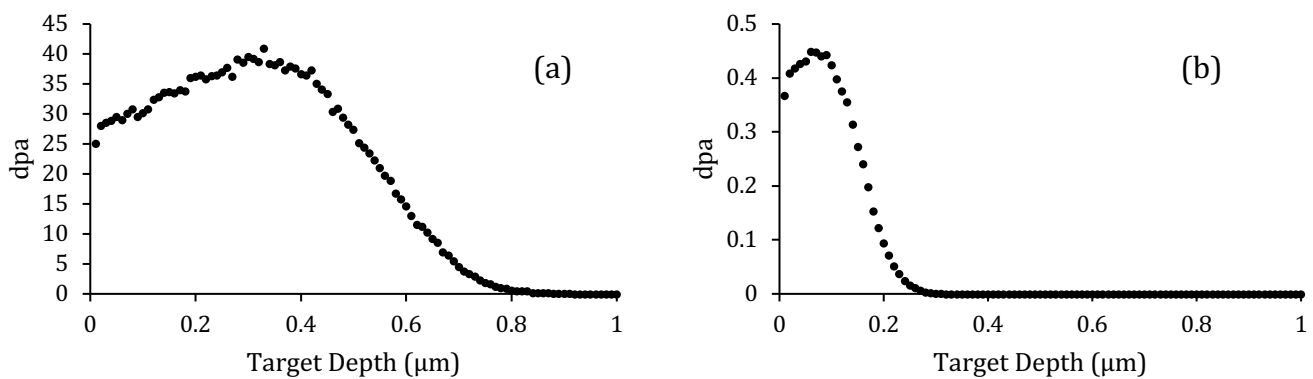

*Figure 2:* SRIM-Predicted implantation profiles for a) 2 MeV $Xe^+$ with a fluence of $1\times10^{16}$ ions/cm$^2$ (showing maximum damage of 40 dpa at 300 nm and total damage depth of 800 nm) and b) 600 keV $Xe^{2+}$ with a fluence of $9\times10^{13}$ ions/cm$^2$ (showing maximum damage of 0.45 dpa at 100 nm and total damage depth of 300 nm).

### 3.3: Superconductivity

In order to assess the degradation in superconducting properties caused by the Xe$^+$ implantation, the critical current density as a function of applied magnetic field ($J_c(B_{app})$) was measured at 4.2 K, before and after implantation. A standard magnetic method was employed. The tape sample (ex-situ implanted with 2MeV Xe ions to a fluence of 1x10$^{16}$ ions/cm$^2$) was cooled in a field value of –1.5 T, which was subsequently swept up to 7 T and back down to –1.5 T, measuring the induced magnetic moment of the sample ($m$) at 0.1 T intervals. Hysteresis loops are generated (as shown in Figure 3 (a)) due to circulating persistent currents in the sample. Employing Bean's model (*23, 24*) $J_c$ is calculated from the hysteresis width ($\Delta m$) at any value of $B_{app}$ by:

$$J_c(B_{app}) = \frac{3}{2\pi R^3 t} \Delta m(B_{app}),$$

where $R$ is the sample radius (1.5 mm), and $t$ is the thickness of the superconducting film (nominally 1 μm). The resultant $J_c(B_{app})$ data are presented in Figure 7 (b) and summarized in Table 1. The secondary axis provides the corresponding wire $I_c$ values (shown in Figure 3b) obtained by multiplying $J_c$ by the width and thickness of the tape. As can be seen, the wire's current carrying ability at 4.2 K is reduced by a factor of 10. This is accompanied by a large reduction in the critical temperature ($T_c$) from 90 K to 80 K. To make a full assessment of the nature and extent of the damage imparted by Xe$^+$ implantation would require detailed measurements to be made over a range of temperatures and field angles, along with measurements of the upper critical field $B_{c2}$. Nevertheless, the measurements made here demonstrate a catastrophic performance reduction that would almost certainly render the wire un-usable in fusion applications at these levels of damage.

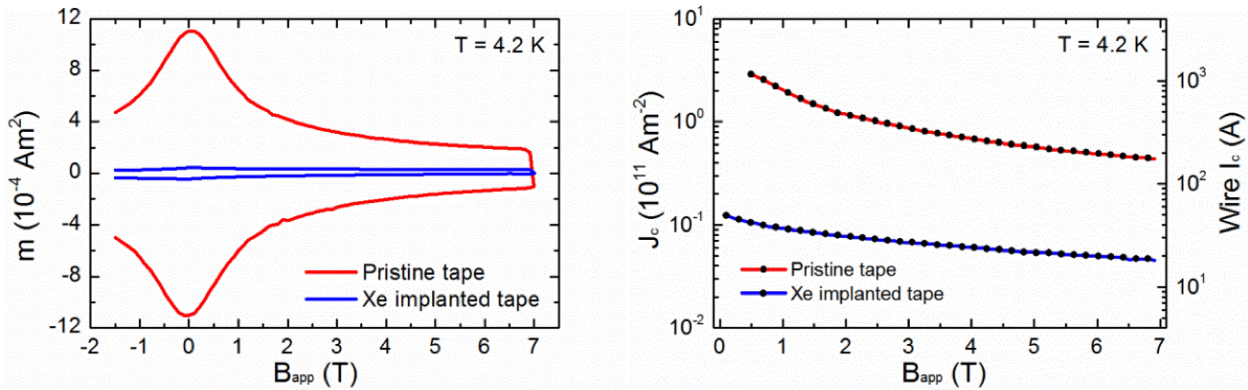

*Figure 3:* SQUID measurements on the tape before and after Xe implantation. (a) Magnetic hysteresis loops measured at 4.2 K for the 2G-HTS tape, before and after Xe implantation. The Cu and Ag overlayers were chemically removed from both samples prior to testing. (b) Critical current density as a function of magnetic field, calculated from the hysteresis widths of the data in (a), employing Bean's model. The secondary axis shows the equivalent critical current values expected for the 4 mm wide tape

*Table 1:* Summary of the superconducting critical temperature in different applied fields and how it varied after ion implantation.

| B[T] | T$_c$ [K] Etched Tape | T$_c$ [K] Xe Implanted Tape |
|---|---|---|
| 0.01 | 90 | 80 |
| 2.00 | 83 | 66 |
| 4.00 | 78 | 55 |

### 3.4: XRD characterisation of implanted bulk and tape samples

X-ray diffractograms of the bulk single crystal (SX) sample and of the etched, implanted Superpower tape sample (Figure 4) confirm the presence of Y-123 and Y-211 phases in the bulk sample and Y-123, γ Ni and MgO in the tape sample. The lattice parameters of the <c> axis in the Y-123 phase were determined to be 11.68 Å (bulk) and 11.74 Å (tape), the implications of lattice parameter and its effect on oxygen content will be elaborated on in the discussion section. In the implanted bulk sample, an amorphous halo was identified in the 20–40° range, which was not present before ion implantation, as shown in Figure 4. This shows that the ion implantation amorphised the bulk sample (to some extent). In the X-ray diffractogram of the implanted tape sample, there was no clear amorphous halo, however, there was a reduction in the intensity of the highest Y-123 peak, relative to the highest nickel substrate peak (going from 33.8, down to 2.5). Since the YBCO layer is 1000 nm thick and the predicted total damage depth for the 2 MeV implantation is 800 nm, it is unlikely that the nickel substrate would have been affected by the ion implantation. Therefore, a reduction in intensity of the Y-123 peaks relative to the nickel peak can be taken as an indicator of loss of crystallinity.

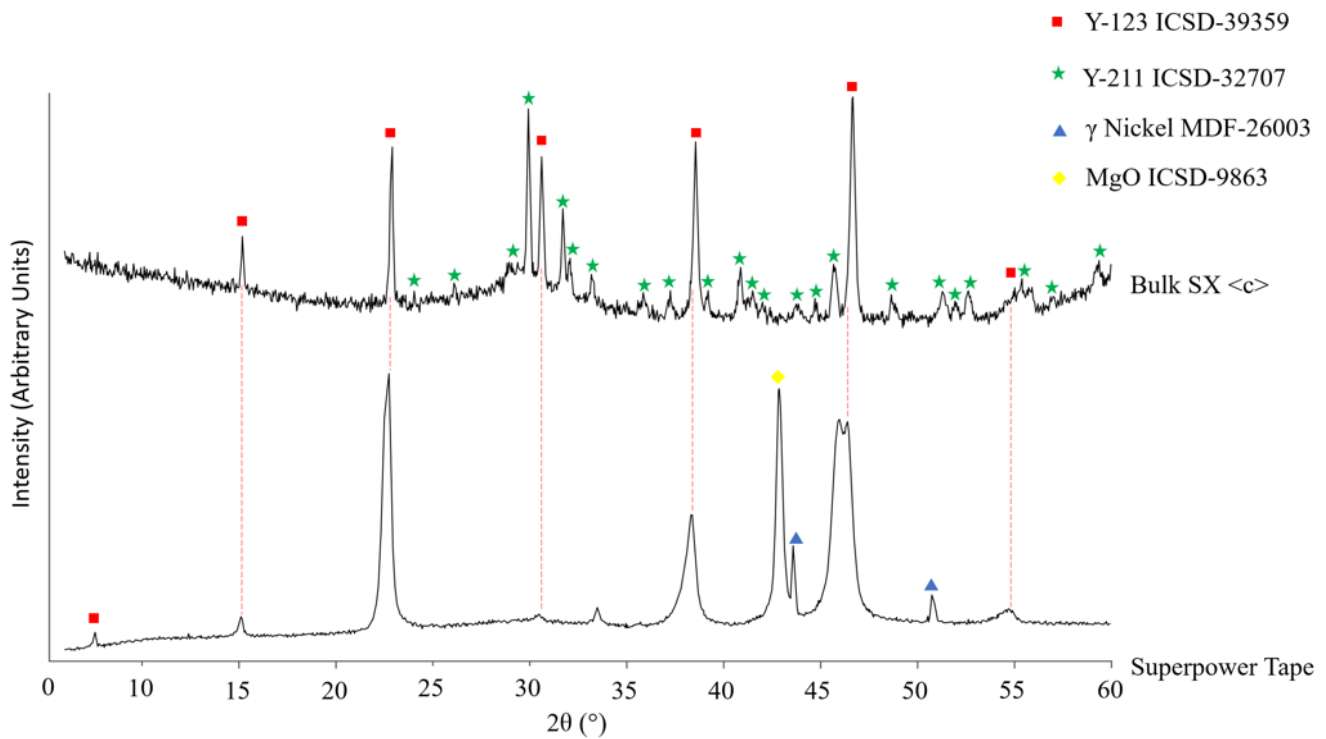

*Figure 4: X-ray diffractograms of the 2 MeV Xe implanted bulk and tape samples ($1 \times 10^{16}$ ions/cm$^2$). Y-123 and Y-211 phases present in the bulk. Y-123, γ-Ni and MgO phases present in the tape. Amorphous halo seen in the bulk between 20-40°.*

### 3.5: Microstructural characterisation of the irradiated sample

The SRIM-predicted damage depths were confirmed by TEM on the bulk sample. Figure 5 (a) shows an overview of the whole TEM sample, as a collage of TEM bright-field micrographs. The surface damage layer is shown to extend to a depth of ~900 nm into the sample. In bright-field TEM micrographs, the damaged layer has homogeneous contrast, which can be a sign of lack of crystallography (further confirmed in Figure 6 through selected area diffraction). However, when imaged with HAADF-STEM (shown in Figure 5 (b)), the enhanced Z-contrast this technique provides shows that local changes in composition remain where the Y-211 particles used to be.

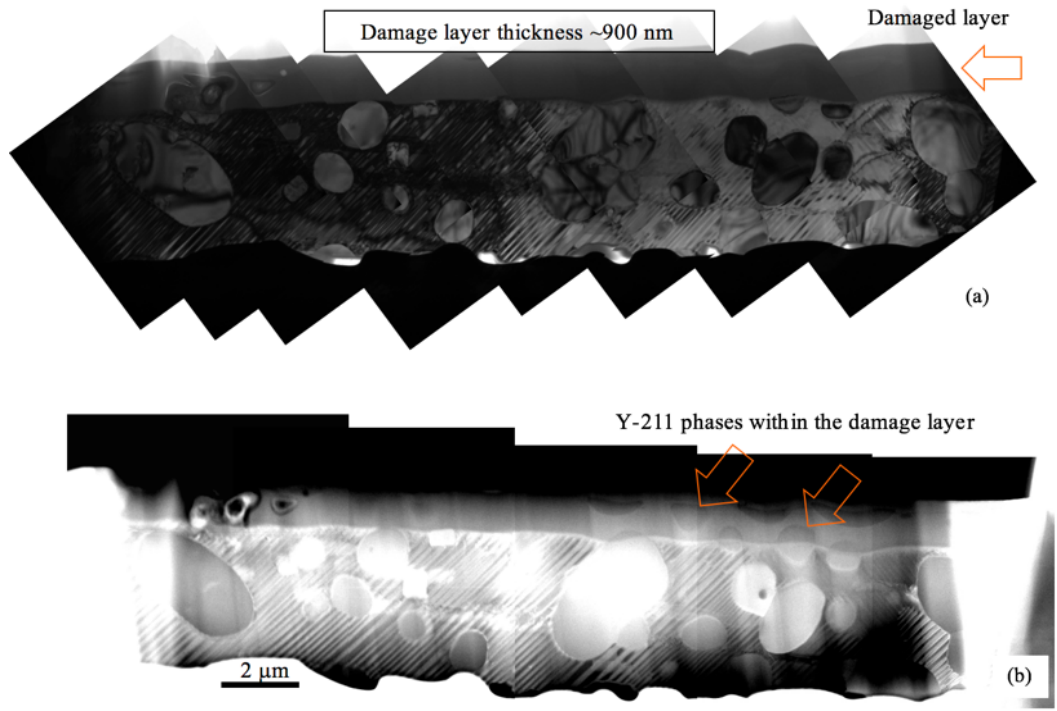

*Figure 5: (a) TEM bright field micrograph of the whole sample, showing the damaged layer on the top surface, several Y-211 particles and the Y-123 matrix. (b) HAADF-STEM micrograph of the same region, taken at the same magnification. This time the improved Z-contrast allows the Y-211 particles in the damaged zone to be clearly seen, despite being amorphous.*

The ion-implanted top surface layer was fully amorphous, as shown through electron diffraction (Figure 6 (d)). Figure 6 (b) shows a TEM bright-field micrograph indicating the regions from which selected area diffraction patterns were acquired, 6 (a) shows an indexed diffraction pattern from the crystalline sub-surface region. Commercial TEM image simulation software JEMS (*25*) was used to generate a theoretical diffraction pattern for Y-123 along the <111> zone axis which was overlaid and proved to match the experimentally acquired pattern in Figure 6 (c). For the simulation, the following database entry was used. From the Inorganic chemistry Structures Database (ICSD): database code ICSD 39359. The c axis is 11.67 Å, which is consistent with XRD measurements. Finally, a diffraction pattern was acquired from the damaged surface region of the sample, which is shown in Figure 6 (d) and shows a fully amorphous structure.

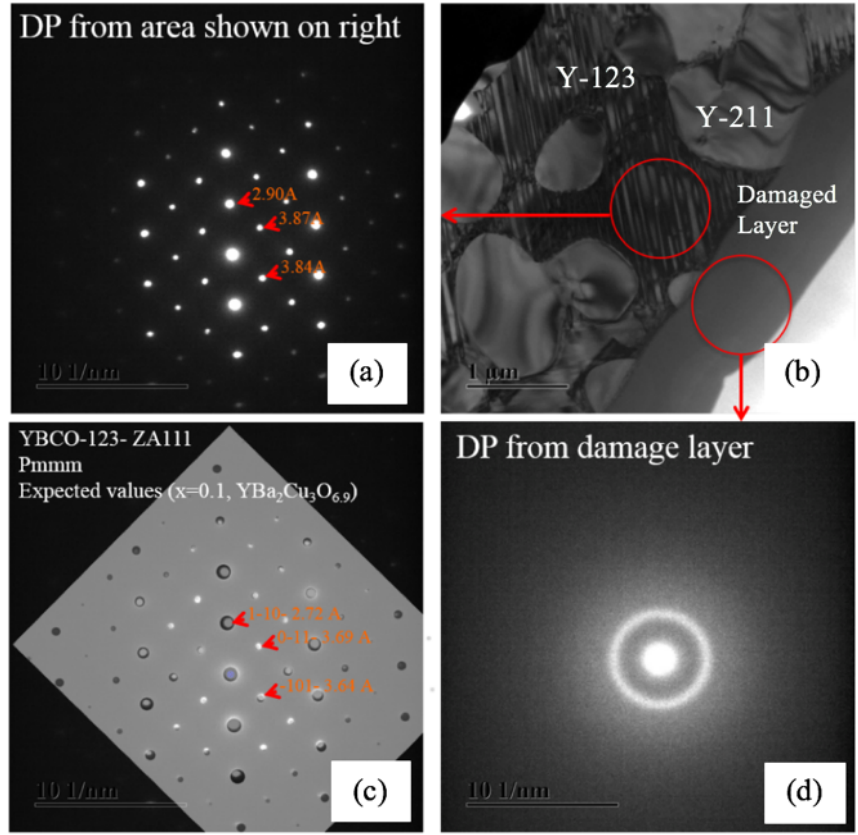

*Figure 6: (a) Diffraction pattern from crystalline region indicated by the arrow in the TEM bright field micrograph shown in (b), and with a theoretical pattern overlaid in (c). (d) shows the diffraction pattern from the fully amorphous damaged surface layer indicated by the arrow.*

Some particles contained within the damage layer were however found to still show crystallographic (channelling) contrast in bright-field micrographs. One such example is shown in Figure 7 (a). When the selected area aperture was used to take a diffraction pattern it became evident that those particles were not fully amorphous. The diffraction pattern, the region from which it was acquired, and the region of the diffraction pattern used for dark field image acquisition are shown in Figure 7 (b). Figure 7 (c) is the corresponding dark-field image. Figure 7 shows the STEM-EDX elemental maps acquired from one such particle, along with the sum EDX signal. The denoising of the EDX maps was performed using principal component analysis, as implemented in the hyperpsy python package (*26*), showing that these crystalline particles contained within the damaged zone consist of barium cerate. Ceria is added in the TSMG process as a Y-211 particle refiner.

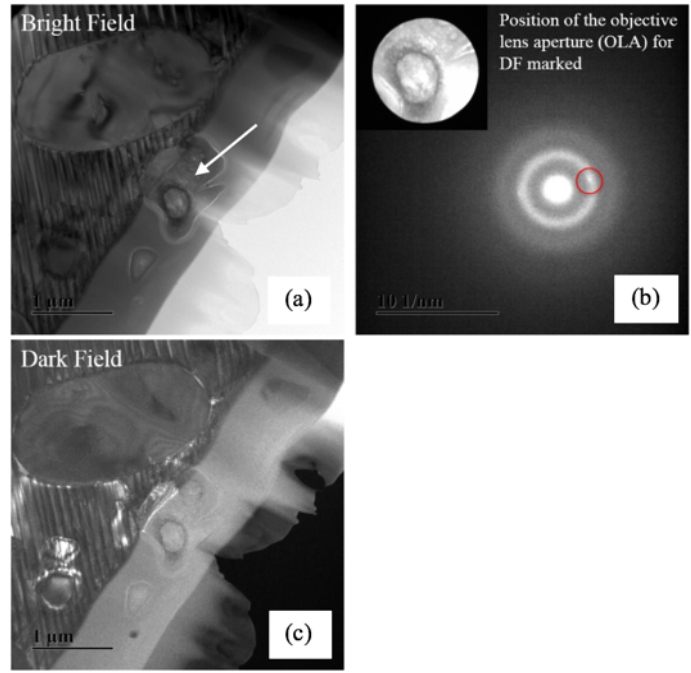

*Figure 7: (a) Bright-field TEM micrograph showing a particle in the damaged layer which still exhibits crystallographic contrast, (b) diffraction pattern and the area it was taken from, showing a partially crystalline structure. (c) dark-field micrograph acquired using the diffracted spot indicated in (b).*

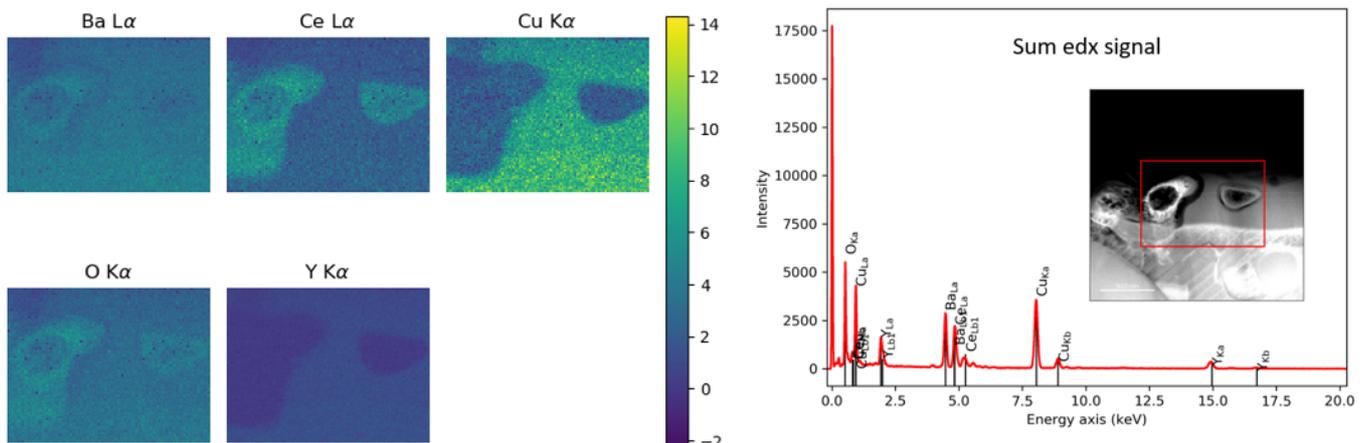

*Figure 8: PCA-deionized STEM-EDX elemental map of the cerium-rich particles which were partially crystalline in the damaged, amorphous layer and sum EDX signal of the same region.*

### 3.6: In-situ TEM irradiation and heating

A TEM sample made of Superpower tape, as shown in Figure 9 (a) along with the results of the in-situ ion irradiation with 600 keV $Xe^{2+}$ at a flux of $6 \times 10^{12}$ ions/cm²/s to a total fluence of $9 \times 10^{13}$ ions/cm². The ion implantation completely amorphised the previously crystalline YBCO sample, as shown by the lack of contrast in the bright field (BF) TEM image in Figure 9 (b), as well as by the fully amorphous SAD (selected area diffraction) pattern associated with it. The sample was then heated in 100°C steps and BF micrographs were taken (Figure 9 c-f), as well as a video (available as supplementary material). At 600°C, partial recrystallisation is evident through contrast from crystalline grains in the BF-TEM image and crystalline spots appearing in the diffraction pattern of Figure 9 (g). At higher temperatures the grains continued to grow in size until the material began to lose structural integrity at 900°C (Figure 9 (h-j)).

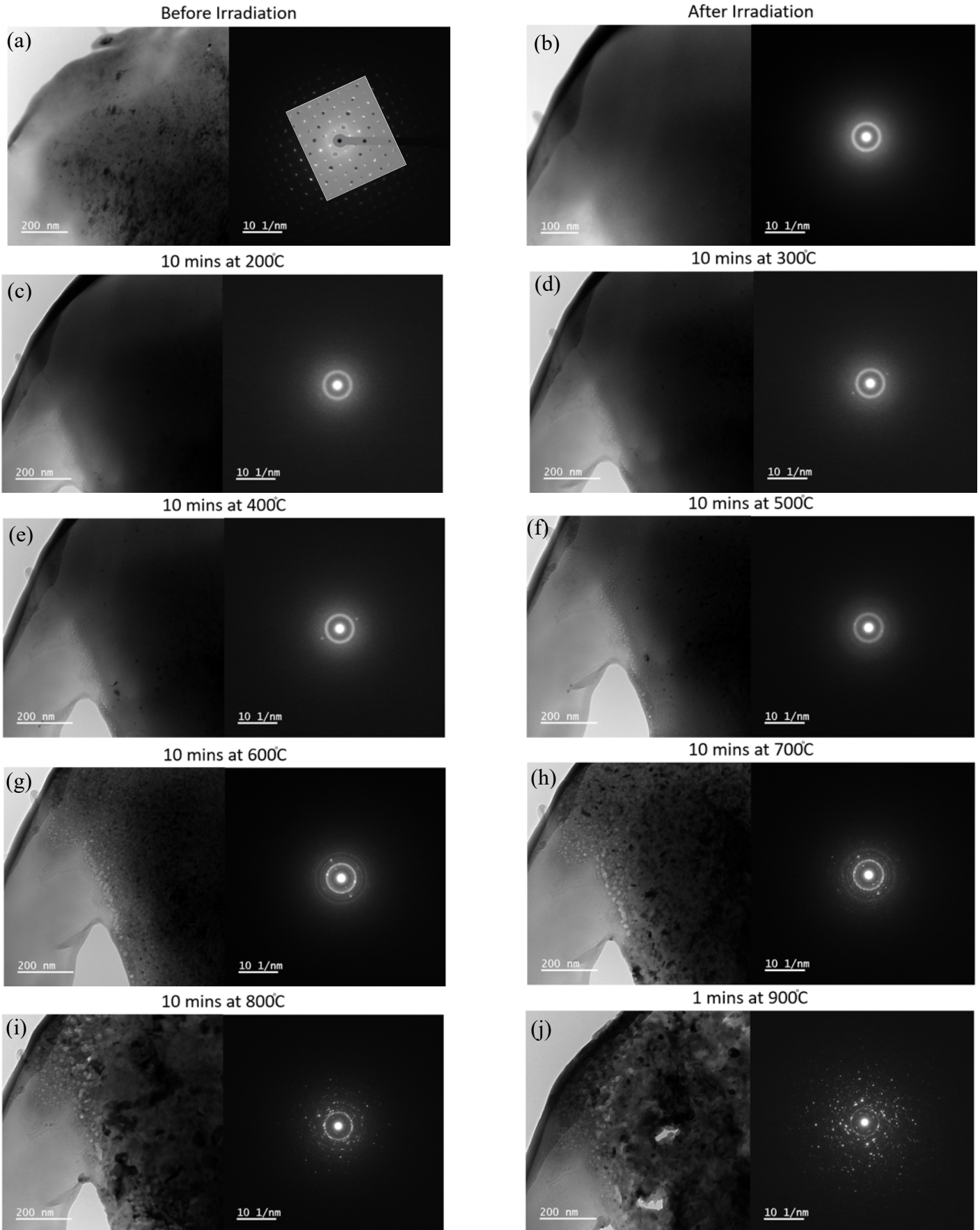

*Figure 9: TEM-BF images and SAD patterns of the etched Superpower tape (a) before (with overlayed simulated pattern of the 111 zone axis, Pmmm structure) and (b) after ion implantation (600 keV $Xe^{2+}$ at $9\times10^{13}$ ions/cm$^2$ fluence). Then (c-j) following each of a series of anneals up to 900˚C. (g) Thermally-induced recrystallisation is seen at 600˚C. (j) At higher temperatures the grains grow larger until the material begins to degrade at 900˚C.*

The Azimuthal radial integration and sum of the SAD patterns in Figure 9 were plotted in Figure 10. The radial integrations allow more accurate interpretation of the amorphization and crystallinity observed in Figure 9 by highlighting crystalline peaks which may not have been previously visible in the unprocessed diffraction patterns. The following trend can be observed in the sum-integrated density of the patterns: the pristine area is fully crystalline, then, when irradiated and heated to 200˚C, a mixture of crystalline and amorphous areas can be observed (t=21 minutes from the start of the heating experiment), around ~300˚C the sample is fully amorphous. After 44 minutes (which corresponds to 400˚C) mass loss is observed, likely from a reduction in oxygen content.

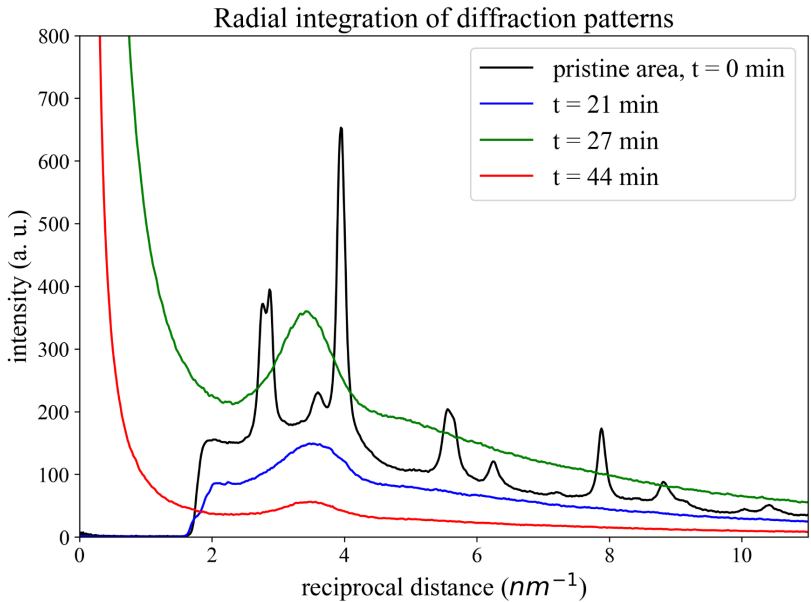

*Figure 10: radial integration of the SAD patterns (pristine, irradiated and heat treated after t=21min (200˚C), t=27 min (~300˚C), t=44 min (400˚C).*

### 3.7: Thermal Recrystallisation through ex-situ Heat Treatments

Eight furnace heat treatments were performed in the 300–800˚C range for durations of 10–1200 mins on the Superpower tape samples (details shown in Table 2). A consistent consequence of heat treatments was an increase in the c lattice parameter, which was due to oxygen loss (*20*). The ion implantation depth was known (from the TEM and SRIM profiles to be <900nm, therefore the height and width of the substrate peaks was deemed unchanged. Peaks shift corrections and normalisation were applied using the γ-nickel peaks. After the corrections, the lattice parameter was calculated (using all the YBCO reflections identified and fitting them with a generated pattern in commercial software CrystalDiffract). Lattice parameter values were given with 2 significant figures, as lab-based XRD cannot give more accurate values, particularly considering the breadth of the peaks. Table 2 shows the lattice parameter before and after heating, though it is noteworthy that some of the measurements error bars partly overlap, eg. sample 8. Furnace heat treatments were performed on ex-situ irradiated samples, with 2 MeV $Xe^{2+}$ ions at a fluence of 1 x $10^{16}$ ions/$cm^2$. Figure 11 shows the comparison between the pristine (unirradiated) tape and Sample 4 (Table 2), which was irradiated then heat treated at 500˚C for 300 minutes. It is noteworthy that in Figure 9, after in-situ ion implantation, the heat treatment caused recrystallisation along random crystal orientations. In Figure 11, it can be seen that when the experiment was performed on bulk samples, only the top 800 nm of the superconducting layer was amorphized by irradiation and the crystal recrystallises with c-axis alignment (90% aligned along the c-axis). This indicates that provided the superconducting layer of YBCO is not fully amorphous, the remaining sections, when heated, could be seeding crystal regrowth with <c> axis alignment.

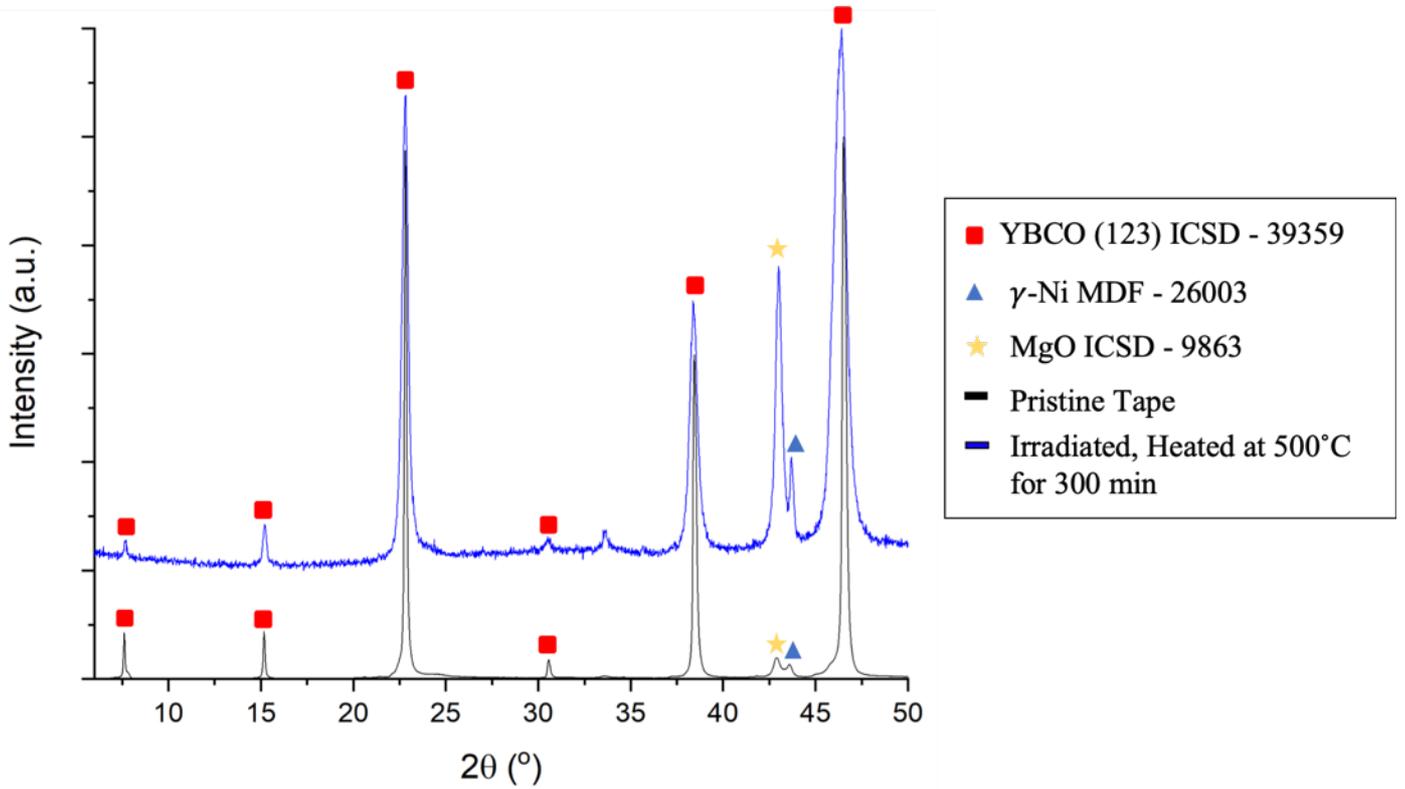

*Figure 11: X-ray diffractograms showing the pristine Superpower tape and comparing it to Sample 4 from Table 2, which was irradiated with 2 MeV Xe at a fluence of $1 \times 10^{16}$ ions/cm$^2$ and then heat treated at 500°C for 300 minutes.*

*Table 2: Heat treatments of etched implanted Superpower tape. Lattice parameters and peak intensity ratios were determined from X-ray diffractograms. Following any heat treatment, the c lattice parameter increased due to oxygen loss.*

| Sample | Temperature (°C) | Duration (minutes) | c Lattice Parameter Before Heating (Å) | c Lattice Parameter After Heating (Å) |
|---|---|---|---|---|
| 1 | 600 | 10 | 11.72±0.01 | 11.76±0.02 |
| 2 | 700 | 10 | 11.72±0.01 | 11.78±0.02 |
| 3 | 800 | 10 | 11.72±0.01 | 11.77±0.02 |
| 4 | 500 | 300 | 11.73±0.01 | 11.76±0.02 |
| 5 | 300 | 480 | 11.73±0.01 | 11.76±0.02 |
| 6 | 300 | 1200 | 11.73±0.01 | 11.76±0.02 |
| 7 | 600 | 120 | 11.73±0.01 | 11.77±0.02 |
| 8 | 500 | 1200 | 11.73±0.01 | 11.76±0.02 |

### 4. Discussion:

The XRD results show that both the unirradiated bulk and tape samples were prepared with the <c> axis normal to the sample surface. Both types of sample contained some form of flux pinning additions, Y-211 particles in the TSMG bulk samples and barium zirconate nanorods in the coated conductor, added to increase critical current density (*27*). Based on previous studies, 167 MeV Xe$^+$ implantation at a fluence exceeding $5 \times 10^{12}$ ions/cm$^2$ is known to fully amorphise YBCO samples (*14*). The present work was performed using 2 MeV Xe ions to a fluence of $1 \times 10^{16}$ ions/cm$^2$ and 600 keV Xe ions to a fluence of $9 \times 10^{13}$ ions/cm$^2$ and both were sufficient to fully amorphise the (previously fully crystalline) perovskite. In the bulk single grain sample, both the Y-123 and Y-211 phases were shown to be fully amorphous after ion implantation. However, some barium cerate particles contained within the damage layer remained crystalline. Ceria has been previously shown to exhibit outstanding phase stability under heavy ion irradiation, for example in the work by Edmondson *et al.*, who implanted it with 3 MeV $Au^+$ ions at $3 \times 10^{15}$ ions/cm$^2$ (*28*). The present observation of crystalline ceria within the damaged zone can however have no influence on the alterations of the superconducting properties of the fully amorphous matrix surrounding it.

The functional properties of amorphous YBCO are well known. It is being studied as it can be produced through deposition or sputtering techniques and has exhibited properties of interest for bolometric applications (*29*). When the material is either fully amorphous or oxygen depleted, YBCO is a semiconductor (*30*). Amorphisation kinetics and recovery mechanisms have not been previously studied in YBCO specifically, though numerous examples of models applied to other ceramics and perovskites can be found, as shown in the review by Weber (*31*). Weber notes that the exact nature of amorphisation is not well defined for most ceramics, however various aspects of it can be described by the models developed. It is also important to note that ceramic amorphisation can occur homogeneously or heterogeneously, and becomes increasingly difficult with increasing temperature and can only occur below a critical temperature (*31*).

Most perovskite ceramics are commonly described as amorphising heterogeneously, by one of several mechanisms, although this can depend on the irradiation conditions used. The most commonly suggested mechanisms, which were proposed for perovskite ceramics such as $ZrSiO_4$ and $SrTiO_3$, are either direct-impact (in-cascade) amorphisation of an individual collision cascade and the local defect accumulation as a result of cascade overlap (*31*). An experimental study performed by HRTEM on $Kr^+$ irradiated YBCO at low fluence, $\sim 10^{11}$ ions/cm$^2$, concludes that the defects observed must be voids $\sim$3-5 nm in diameter, though no comment is made regarding the possibility their observations could correspond instead to either ion tracks from swift heavy ions, or amorphous zones, or just low density regions (*32*). By comparison, in the present work, voids were not observed during ion implantation, however, during the subsequent heat treatment, voids (or potential Xe bubbles) were found to coalesce and grow. A numerical study of ion-induced thermal spikes in YBCO predicted that cascades due to irradiation with 107 MeV $Kr^{17+}$ ions, may cause localized melting (*32*). Their results estimate the characteristic size of the regions at $\sim$50-70 nm, but variable depending on the ions used for implantation (*33*). These findings suggest that amorphisation in YBCO may be heterogeneously nucleated and correlated mainly to the number of direct impacts coupled with defect accumulation due to cascade overlap, although further studies are required to confirm the amorphization mechanisms.

Analytical models to describe this coupled effect were developed by Gibbons (*34*). However, they neglect to account for any material recovery processes. Two main recovery processes reduce the amorphisation rate in ceramics: thermal and irradiation assisted recovery (i.e. the models work at 0 K, and during ion implantation at room temperature some thermal recovery will occur which will not be taken into account by these models). Some perovskites such as $ZrSiO_4$ have been shown to exhibit mainly thermal-assisted recovery (*31*). In order to attempt to model the findings, further studies are needed to determine the amorphous fraction as a function of fluence at room temperature when compared to cryogenic irradiation.

The concept of amorphisation becoming increasingly difficult with increasing temperature and only occurring below a specific critical temperature (*31*) led us to explore the possibility of annealing damage through heat treatments to attempt recovery of the functional properties of the YBCO. If the stoichiometry is unaltered after ion implantation, the crystal structure and therefore the superconductivity could potentially be restored by thermally assisted recovery processes. Thermal recrystallisation could lead to the component in-service lifetime to be significantly extended, reducing the cost and environmental impact. This possibility was tested by our experiments with in-situ TEM implantation and annealing as well as ex-situ furnace heat treatments and XRD. In-situ heating (Figure 9) confirmed that the YBCO can be recrystallised in vacuum at 600°C, going from fully amorphous to random polycrystalline, though it remains unclear whether annealing before the material became fully amorphised could lead to a return to the original structure. Grain growth could be induced subsequently through further increases in temperature (up to 800°C), leading to fewer, larger grains. The work by Tate *et al.* (*15*) suggested that holding the samples for 10s at 820°C in air would allow the recovery of the properties, however, our work showed that at 900°C, the material degraded beyond recovery as evidenced by the structures formed in Figure 9.j. Whether or not any trapped Xe-ions present in the lattice play a role in the degradation remains to be explored.

Oxygen loss in YBCO can be observed as an increase in the orthorhombic unit cell's lattice <c> parameter with decreasing oxygen content, as demonstrated by Benzi et al. (20), who derived a linear relationship described by the equation **(7 – δ) = 75.250 – 5.856c**. Table 3 shows estimates of the oxygen content of the YBCO samples after heat treatments, obtained by processing XRD data from Table 2 using the equation derived by Benzi *et al*. Table 3 compares alterations in lattice parameters after heat treatment of both implanted (lines 1–8) and non-implanted samples (lines A–C,) to ensure that alterations are solely the result of the heat treatment and not the implantation.

Table 3: Determined oxygen stoichiometries of heat-treated samples. Pristine (non-implanted) samples are A–C. Implanted samples are 1–8. *errors for samples 1-8 were calculated across multiple samples of each condition. They could not be caculated for A-C because only one sample was analysed.,

| Sample | Temperature (˚C) | Duration (mins) | Oxygen Stoichiometry Before | Oxygen Stoichiometry After |
|---|---|---|---|---|
| A | 680 | 5 | 6.68±0.04 | 6.40±0.15 |
| B | 900 | 10 | 6.62±0.04 | 6.25±0.15 |
| C | 700 | 300 | 6.68±0.04 | 6.32±0.15 |
| 1 | 600 | 10 | 6.62±0.12 | 6.38±0.23 |
| 2 | 700 | 10 | 6.62±0.12 | 6.27±0.12 |
| 3 | 800 | 10 | 6.62±0.12 | 6.32±0.23 |
| 4 | 500 | 300 | 6.60±0.06 | 6.38±0.23 |
| 5 | 300 | 480 | 6.60±0.06 | 6.38±0.23 |
| 6 | 300 | 1200 | 6.60±0.06 | 6.38±0.23 |
| 7 | 600 | 120 | 6.60±0.06 | 6.32±0.23 |
| 8 | 500 | 1200 | 6.60±0.06 | 6.38±0.23 |

The objective of the heat treatments was to recrystallise the YBCO whilst retaining both oxygen content and superconducting properties. The in-situ implantation and heating, as well as heat treatments of the pristine Superpower tape, provided a guideline for temperature and duration of the heat treatments performed on the implanted etched tape samples (shown in Table 3). The values shown in Table 3 suggest that neither temperature nor duration have a significant impact on the final oxygen stoichiometry, which is known to correlate closely with superconductivity (*35*).

### 5. Conclusions:
The microstructure and electrical properties of YBCO superconducting tape and bulk samples were assessed before and after ion implantation. The present study proved that:

1. ***Xe-ion implantation at an energy of 600 keV to a fluence of 1x10$^{13}$ ions/cm$^2$*** amorphised the TEM lamellae (the whole depth) during in-situ TEM ion implantation experiments. Subsequent in-situ heating of the lamellae caused recrystallisation at around ~600°C when samples were held for 10 min and melting/complete degradation of the samples at 900°C.
2. ***Xe-ion implantation at an energy of 2 MeV to a fluence of 1x10$^{16}$ ions/cm$^2$*** reduced the superconducting temperature by 10 K and cut the critical current density down to 1/10$^{th}$ of its former value. The fluence and energy selected were predicted to produce damage to a depth of 800–900 nm below the surface of the samples, using SRIM software. TEM characterisation revealed that the specimen was fully amorphized to a depth of 900 nm after ion implantation. Of the original 1μm thick YBCO tape, only 1/10$^{th}$ remained crystalline and superconducting.
3. Heat treatments showed that recrystallisation can be achieved around 600˚C, however the heavily textured <c> axis grain orientation which was present before irradiation cannot be recovered unless the superconductive tape was not fully amorphised. The lattice parameter shows some oxygen loss after heat treatment. Future studies annealing from an earlier stage (prior to full amorphization) to recover the properties would be of interest.

**Acknowledgements:**
Funding is acknowledged from the UK Engineering and Physical Sciences Research Council (EPSRC) under grants EP/K029770/1 and EP/M028283/1. PDE also acknowledges support from the U. S. Department of Energy, Office of Science, Fusion Energy Sciences as well as EPSRC funding under grant EP/K030043/1. We would also like to thank industrial collaborator SuperPower Inc., the Bulk Superconductivity Group at the University of Cambridge and Prof Chris Grovenor, Dr Susie Speller and Dr Paul Bagot at the University of Oxford for sample provision.

Part of this work has been carried out within the framework of the EUROfusion Consortium and has received funding from the Euratom research and training programme 2014-2018 and 2019-2020 under grant agreement No 633053. The views and opinions expressed herein do not necessarily reflect those of the European Commission.